\documentclass{rspublic}
\usepackage{natbib}
\usepackage{amssymb,graphicx,amsmath}
\usepackage{color}
\bibpunct{(}{)}{;}{a}{}{,}
\begin{document}

\title{Synchronization and entrainment of coupled circadian oscillators}
\author{N. Komin, A.C. Murza, E. Hern\'{a}ndez-Garc\'{\i}a, R. Toral}
\affiliation{IFISC (Instituto de F\'{\i}sica Interdisciplinar y
Sistemas Complejos), CSIC-UIB, Campus UIB, E-07122 Palma de
Mallorca, Spain}

\label{firstpage}
\maketitle
\begin{abstract}{
Circadian oscillations; quenched noise; noise-induced
oscillators death; modified Goodwin model; noise-induced
synchronization.}

Circadian rhythms in mammals are controlled by the neurons located in the suprachiasmatic nucleus of the hypothalamus. In physiological conditions, the system of neurons is very efficiently entrained by the 24-hour light-dark cycle. Most of the studies carried out so far emphasize the crucial role of the periodicity imposed by the light dark cycle in neuronal synchronization. Nevertheless, heterogeneity as a natural and permanent ingredient of these cellular interactions is seemingly to play a major role in these biochemical processes. In this paper we use a model that considers the neurons of the suprachiasmatic nucleus as chemically-coupled modified Goodwin oscillators, and introduce non-negligible heterogeneity in the periods of all neurons in the form of quenched noise. The system response to the light-dark  cycle periodicity is studied as a function of the interneuronal coupling strength, external forcing amplitude and neuronal heterogeneity. Our results indicate that the right amount of heterogeneity helps the extended system to respond globally in a more coherent way to the external forcing. Our proposed mechanism for neuronal synchronization under external periodic forcing is based on heterogeneity-induced oscillators death, damped oscillators being more entrainable by the external forcing than the self-oscillating neurons with different periods.
\end{abstract}

\section{Introduction}
\label{introduction}

Circadian rhythms are light-dark dependent cycles of roughly 24 hours present in the biochemical and physiological processes of many living entities~\citep{RW02}. In mammals the main mediator between the light-dark periodicity and the biological rhythms is formed by two interconnected suprachiasmatic nuclei (SCN), located in the hypothalamus. These nuclei form the so called ``circadian pacemaker" and contain about $10.000$ neurons each~\citep{RW02,MSL02}.

The main property of the SCN is that their activity displays self-sustained oscillations in synchrony with the external forcing imposed by the light-dark cycle. The exact mechanism leading to this behavior has been the subject of intense research. It has been shown that, when taken individually, neurons produce oscillations with a constant period  ranging from 20 to 28 hours~\citep{Honma2004,WLM95}. The oscillatory behavior originates in a regulatory circuit with a negative feedback loop. The relevant question is how this individual oscillatory behavior translates into common, global, oscillations of the SCN activity synchronized with the external light stimulus.

It has been shown that the origin of the oscillatory activity of the circadian pacemaker at the global level resides on the interaction between the SCN neurons. Coupling between cells in the SCN is achieved partly by neurotransmitters~\citep{Honma2004,HH04} and it is by means of those neurotransmitters that external forcing by light influences the neuronal synchronization.  For example the vasoactive intestinal polypeptide (VIP) has been shown to be necessary in mediating both the periodicity and the internal synchrony of mammalian clock neurons \citep{Shen2000,Aton2005NatNeuro,Maywood2006}. Therefore, a model of coupled and forced neurons appears quite naturally as responsible for the circadian rhythms. Along these lines, an interesting mechanism has been put forward recently by~\citet{DidierGonze07012005} and by~\citet{BJ07}. They proposed that synchronization to the external forcing is facilitated by the fact that interneuronal coupling transforms SCN into damped oscillators which can then be easily entrained.

In this paper we show that the presence of some level of
heterogeneity or dispersion in the intrinsic periods of the
oscillators~\citep{SAV03,HAN04} can improve the response of
the coupled neuronal system to the external light-dark forcing.
The proposed mechanism for the improvement of the neuronal
synchronization under external periodic forcing bears some
similarities with the one proposed in~\citep{DidierGonze07012005,BJ07} in the
sense that the oscillators are brought to a regime of
oscillator death \citep{Ermentrout1990219,MS:1990}, but in our case this
regime is induced by the presence of heterogeneity. Once this
regime has been reached, the damped oscillators are more
entrainable by the external forcing than the self-oscillating
neurons with different periods, or the synchronized oscillatory
state which appears in the strong coupling regime but with a
period larger than the individual neuronal periods.

To be more specific, we will assume that the periods of the
individual neurons are random variables drawn from a normal
distribution. We will then analyze the global response of the
system to the light-dark  cycle periodicity as a function of
the interneuronal coupling strength, external  forcing
amplitude and neuronal heterogeneity. We show that the
presence of the right amount of dispersion in the periods of
the neurons can indeed enhance the synchronization to the
external forcing.

Period dispersion arises as a consequence of the cellular heterogeneity at the biochemical level, which is an experimentally well observed fact~\citep{Aton2005,Honma2004}. It can act in either physiological or pathological conditions. An example of the latter is the diversification of antigenic baggage present in tumor cells that makes them more difficult to be recognized and captured by the defense mechanisms and therefore more prone to migrate and develop metastasis~\citep{Gonzalez2002}. Our results show that some level of disorder can be of help when synchronizing neuronal activity to the external forcing. Although counterintuitive, it has been unambiguously shown that the addition of various forms of disorder can improve the order in the output of a large variety of nonlinear systems. For example, the mechanism of {\sl stochastic resonance} \citep{Gammaitoni1998,HM} shows that the response of a bistable system to a weak signal can be optimally amplified by the presence of an intermediate level of dynamical noise. Stochastic resonance is not a rare phenomenon; it has been repeatedly shown to be relevant in physical and biological systems described by nonlinear dynamical equations~\citep{Gammaitoni1998,HM}. In large systems with many coupled elements, noise is responsible for a large variety of ordering effects, such as pattern formation,  phase transitions, phase separation, spatiotemporal stochastic resonance, noise-sustained structures, doubly stochastic resonance, amongst many  others~\citep{GS:99}. All these examples have in common that some sort of order at the macroscopic level appears only in the presence of the right amount of noise or disorder at the microscopic level. Furthermore, it has been proven that noise may play a constructive role in nonlinear systems, by enhancing coherent (periodic) behavior near bifurcations and phase transitions~\citep{Neiman97,PK:1997}. In this paper we introduce non-negligible random heterogeneity into the periods of all neurons, so-called quenched noise. Numerical simulations suggest (data not shown) that the results are valid as well when the quenched noise is introduced into the model parameters. A different approach is the consideration of intracellular stochastic variability due to low molecule numbers~\citep{Forger2005} or both variability and heterogeneity.

Close to our work is the study by~\citet{UHI02}, where the effect of fluctuations in neuron parameter values is assessed and it is shown that the coupled system is relatively robust to noise. Previous theoretical studies have addressed the effect of noise on genetic oscillators~\citep{VO01,Steuer2003,VO05}, and some have proposed an ordering influence of noise on circadian clocks at the single cell level in cases where neither light intensity nor coupling strength by themselves can synchronize the system. Collective phenomena induced by heterogeneity in autonomous, non-forced systems, has also been discussed in the literature. For example~\citep{deVries2001} and~\citep{Cartwright2000} have shown that collective bursting or firing can appear in excitable systems and a general theory of the role of heterogeneity in those systems has been developed by~\citep{tessone2007a}. In this paper, we refer to the collective response in systems of non-linear oscillators subjected to the action of an external forcing representing the day-light cycle. 

The paper is organized as follows. In section \ref{sec:Model} we will describe in detail the model of circadian oscillators and the methods we use. It is a coupled extension of the original Goodwin oscillator~\citep{Goodwin1965} as developed by~\citet{DidierGonze07012005}. In section \ref{sec:Results} we analyze the system response to the periodic external forcing, as a function of the external forcing amplitude, coupling strength and neuronal diversity or heterogeneity. By simulating numerically the governing differential equations we identify the range of these parameters for which the extended system oscillates in synchrony and entrained to the external light period. Section \ref{sec:Discussion} describes the mechanism through which the neuronal heterogeneity favors the synchronization with the external forcing and analyzes the combined influence of the coupling strength, neuronal heterogeneity and light amplitude on the stability of the linearized system of coupled oscillators. We show that a mean variable in this model exhibits a transition from a rhythmic to an arrhythmic dynamics (the so-called oscillator death \citep{Ermentrout1990219,MS:1990}). Concluding remarks are found in section~\ref{sec:Conclusion}.

\section{Model and methods}
\label{sec:Model}

\subsection{The circadian pacemaker}
As stated in the introduction, our aim is to consider the role
that the heterogeneity in the population of neurons plays in
the global response of the SCN to an external oscillating
stimulus. To this end, we consider an ensemble of coupled
neurons subject to a periodic forcing. Each of the neurons,
when uncoupled from the others and from the external stimulus,
acts as an oscillator with an intrinsic period. Heterogeneity
is considered insofar the individual periods are not identical,
but show some degree of dispersion around a mean value.  For
each one of the neurons in the SCN we use a four-variable model
proposed by Gonze et al.~\citep{DidierGonze07012005}, which is based originally
on the Goodwin oscillator~\citep{Goodwin1965}, to describe circadian
oscillations in single cells. The variables of the model are as
follows: The clock gene {\sl mRNA} ($X$) produces a clock
protein ($Y$), which activates a transcriptional inhibitor
($Z$) and this in turn inhibits the transcription of the clock
gene, closing a negative feedback loop. The {\sl mRNA} $X$ also
excites the production of neurotransmitter $V$, which in the
coupled system will be then the responsible of an additional
positive feedback loop. In order to overcome the high Hill
coefficients required for self-oscillations, Gonze et al.
replaced the linear degradation by nonlinear Michaelis-Menten
terms. This leads to the system of equations:
\begin{eqnarray}
\label{eq:Goodwin1} \frac{dX}{dt}&=& \nu_1\frac{K_1^{4}}{K_1^4+Z^4}-\nu_{2}\frac{X}{K_2+X},\\
\label{eq:Goodwin2} \frac{dY}{dt}&=& k_3X-\nu_{4}\frac{Y}{K_4+Y},\\
\label{eq:Goodwin3} \frac{dZ}{dt}&=& k_5Y-\nu_{6}\frac{Z}{K_6+Z},\\
\label{eq:Goodwin4} \frac{dV}{dt}&=& k_7X-\nu_{8}\frac{V}{K_8+V},
\end{eqnarray}
which, depending on parameters, might produce oscillations in a
stable limit cycle. Using the values $\nu_1=0.7$~nM/h,
$\nu_2=\nu_4=\nu_6=0.35$~nM/h, $\nu_8=1$~nM/h,
$K_1=K_2=K_4=K_6=K_8=1$~nM, $k_3=k_5=0.7$/h, $k_7=0.35$/h, the
period of the limit cycle oscillations is $T=23.5$~h.

For the complete model, we take $N$ neuronal oscillators, each one of them described by four variables $(X_i,Y_i,Z_i,V_i)$, $i=1,\dots,N$, satisfying the above evolution equations. Heterogeneity in the intrinsic periods is introduced by multiplying the left-hand-side of each one of the equations (\ref{eq:Goodwin1}--\ref{eq:Goodwin4}) by a scale factor $\tau_i$. Hence, the intrinsic period $T_i$ of the isolated neuron $i$ is $\tau_i T$. The numbers $\tau_i$ are independently taken from a normal random distribution of mean $1$ and standard deviation $\sigma$. Since the periods must be positive, in the numerical simulations we have explicitly checked that, for the values of $\sigma$ considered later, $\tau_i$ never takes a negative value, which would be unacceptable. The standard deviation $\sigma$ will be taken as a measure of the diversity. A value of $\sigma=0.1$ for example corresponds to a standard deviation of $10\%$ in the individual periods of the uncoupled neurons, close to the observed variation of periods between $20$ and $28$ hours.

Two additional factors influence the dynamics of single cell oscillations: forcing by light and intercellular coupling. Both are assumed to act independently from the negative feedback loop and are added as independent terms in the transcription rate of $X$~\citep{DidierGonze07012005}. Light is incorporated through a periodic time-dependent function $L(t)$, which can be realized in various forms. In the majority of the presented results we have used a sinusoidal signal, $L(t)=\displaystyle\frac{L_0}{2}\left(1+\sin \omega t\right)$. In some cases, for comparison and to simulate different day lengths, we have used a square wave $ L(t)=\left\{
	\begin{array}{cl} 
		L_0, & \mbox{if }(t\,{\rm mod}\,24h) < t_{light}\\ 
		0, & \mbox{otherwise} 
	\end{array}
\right. $. In both ways the signal oscillates between the values $L(t)=0$ and $L(t)=L_0$ with a period $2\pi/\omega=24h$.

Coupling between the neurons is assumed to depend on the concentration $F$ of the synchronizing factor (the neurotransmitter) in the extracellular medium, which builds-up by contributions from all neurons. Under a fast transmission hypothesis, the extracellular concentration is assumed to equilibrate to the average, mean-field, cellular neurotransmitter concentration, $F=\frac{1}{N}\sum_{i=1}^{N}V_{i}$.  The resulting model is: 
\begin{eqnarray}
\label{eq:Gonze1} \tau_i\frac{dX_i}{dt}&=& \nu_1\frac{K_1^4}{K_1^4+Z_i^4}-\nu_{2}\frac{X_i}{K_2+X_i}+\nu_{c}\frac{KF}{K_c+KF}+L(t)\\
\label{eq:Gonze2} \tau_i\frac{dY_i}{dt}&=&k_3X_i-\nu_{4}\frac{Y_i}{K_4+Y_i},\\
\label{eq:Gonze3} \tau_i\frac{dZ_i}{dt}&=&k_5Y_i-\nu_{6}\frac{Z_i}{K_6+Z_i},\\
\label{eq:Gonze4} \tau_i\frac{dV_i}{dt}&=&k_7X_i-\nu_{8}\frac{V_i}{K_8+V_i},\\
\label{eq:Gonze5} F&=&\frac{1}{N}\sum_{i=1}^{N}V_{i},
\end{eqnarray}
with $\nu_c=0.4$~nM/h, $K_c=1$~nM.

There is experimental evidence supporting the assumption of a chemical (rather than electrical) mechanism of inter-cell communication among SCN neurons as a synchronization factor and, in fact, mechanisms other than neurotransmitters or electrical coupling for the SCN communication have been suggested (e.g. by \citet{VanDenPol1993}). Furthermore, more realistic modeling which takes into account all variables known to participate of the negative feedback loop has been introduced. These models may include up to $10$ variables and corresponding equations for each single cell~\citep{{BJ07}}.  

It seems, however, that in order to get understanding of the SCN dynamics, a sufficient tool is the $4$ variable model described above. In fact, the synchronization of damped oscillators is independent from the particular intracellular model used and as discussed by~\citep{BJ07}, this system, the model developed by~\citep{Leloup2003}, and other simple negative feedback oscillators have similar synchronization properties. 
In this paper we have decided to use the simpler 4-variable model although most of our results are also valid in the more complex 10-variable model.

A model close to (\ref{eq:Gonze1}--\ref{eq:Gonze5}) has been used by \citet{Ullner2009}, where the authors investigate how the interplay between fluctuations of constant light and intercellular coupling affects the dynamics of the collective rhythm in a large ensemble of non-identical, globally coupled oscillators. In their case, however, an inverse dependence of the cell-cell coupling strength on the light intensity was implemented, in such a way that the larger the light intensity the weaker the coupling.

\subsection{Measures of synchrony and entrainment}
Due both to the effect of coupling and of forcing, the neurons
might synchronize their oscillations. There are several
possible measures of how good this synchronization is. In this
paper, the interneuronal synchronization will be quantified by
the parameter of synchrony $\rho$,
defined as
\begin{equation}\label{RG}
\rho=\sqrt{1-\left\langle \frac{\sum_{i=1}^N[V_i(t)-F(t)]^2}{\sum_{i=1}^NV_i(t)^2} \right\rangle}=
\sqrt{\left\langle \frac{F(t)^2}{\frac{1}{N}\sum_{i=1}^NV_i(t)^2} \right\rangle},
\end{equation}
where $\langle \dots\rangle$ denotes a time average in the
long-time asymptotic state. The parameter $\rho$ varies between
a value close to $0$ (no synchronization) and $1$ (perfect
synchronization, with all neurons in phase, $V_i(t)=V_j(t),
\forall i,j$). It is important to note that even if the neurons
synchronize perfectly their oscillations, the period of those
oscillations does not necessarily coincide with the mean period
$T$ of the individual oscillators or with the period
$2\pi/\omega$ of the external forcing. In fact, in the unforced
(no light) case, the period of the common oscillations (for the
set of parameters given before and a dispersion of
$\sigma=0.05$ and coupling $K=0.5$) is approximately equal to $26.5$~h whereas the
period of the forcing is $2\pi/\omega=24$~h and the mean period
of the individual uncoupled oscillators is $T=23.5$~h
~\citep{DidierGonze07012005}.

Besides the previous measure of synchronization amongst the
oscillators, we are also concerned about the quality of the
global response of the neuronal ensemble to the external
forcing $L(t)$. A suitable measure of this response can be
defined using the average gene concentration,
\begin{equation}
    {\bf X}(t)=\frac{1}{N}\sum_{i=1}^N X_i(t),
\end{equation}
and computing the so-called spectral amplification factor $R$~\citep{Gammaitoni1998},
\begin{equation}
    \label{eq:R}
    R=\frac{4}{L_0^2}\left|\langle e^{-i\omega t} {\bf X}(t)\rangle\right|^2.
\end{equation}
$R$ is nothing but the normalized amplitude of the Fourier
component at the forcing frequency $\omega$ of the time series
$\bf{X}(t)$. We will show that, under some circumstances, the
response $R$ will increase with the intrinsic diversity
$\sigma$ and that the period of the oscillations at the global level coincides with that of the external forcing, these being the main results of this paper.

\section{Results}
\label{sec:Results}

The synchronization properties of the set of circadian oscillators is influenced by the amplitude of the external forcing $L_0$, the coupling strength $K$ and the diversity in the individual periods $\sigma$. The role of the first two has been studied in~\citep{BJ07,BJ04,DidierGonze07012005}. In this section we focus on the heterogeneity of neuronal periods and analyze the combined influence of $L_0$, $K$ and $\sigma$ on the different parameters quantifying interneuronal synchronization and response to the forcing.

\begin{figure}
    \begin{center}
      \includegraphics[width=1.\textwidth]{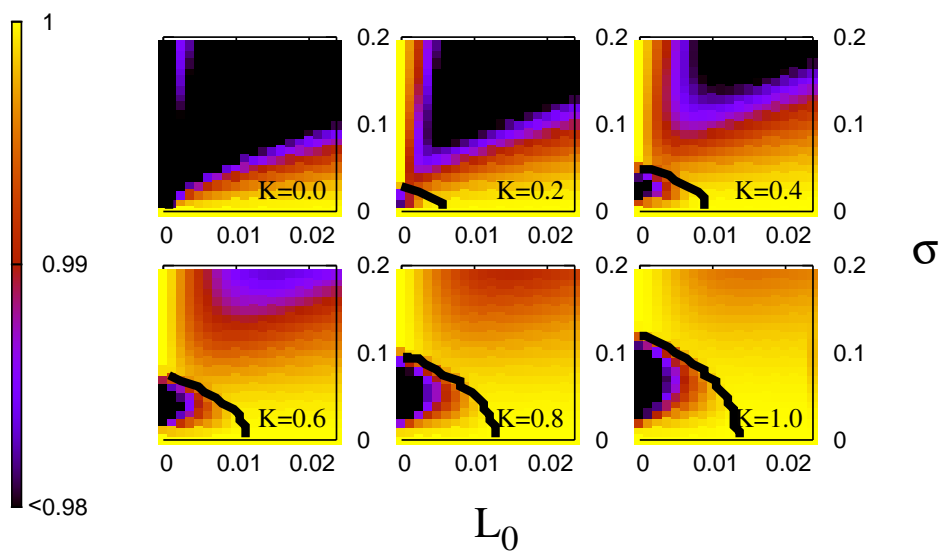}
    \end{center}
    \caption{Synchrony order parameter $\rho$ (see Eq.~(\ref{RG})). Values are coded in colour levels, and displayed as a function of $L_0$ and $\sigma$ for several values of $K$. Data from numerical simulations of $N=1000$ neurons with dynamics ruled by Eqs.~(\ref{eq:Gonze1}--\ref{eq:Gonze5}). Synchrony among the neurons (yellow region) is favored by strong or very weak light intensity $L_0$, low diversity $\sigma$ and large coupling $K$. The thick black line is the linear stability limit discussed in section~\ref{sec:Discussion} (see also Fig.~\ref{fig:contourEVS}).}
    \label{fig:contourRho}
\end{figure}

\begin{figure}
  \begin{center}
  \includegraphics[width=8cm]{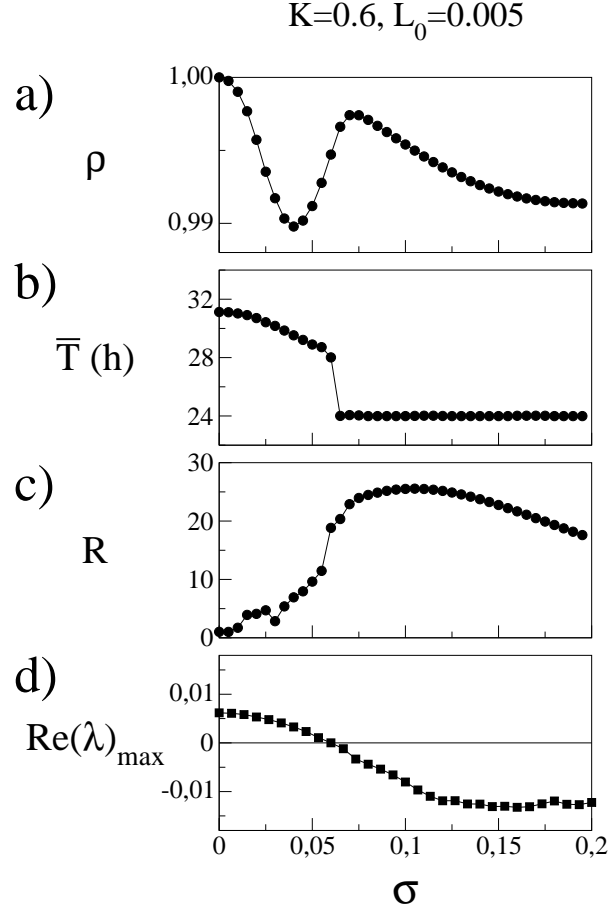}
  \end{center}
  \caption{Main parameters used for characterizing the synchronization of circadian oscillators as a function of the variance $\sigma$. (a) the synchrony parameter $\rho$; (b) the mean $\bar T$ of the individual periods $T_i$;  (c) the response order parameter $R$; (d) the maximum real part of the eigenvalues of the linearized system.}
  \label{fig:orderParameters}
\end{figure}

\begin{figure}
  \begin{center}
  \includegraphics[width=8cm,angle=270]{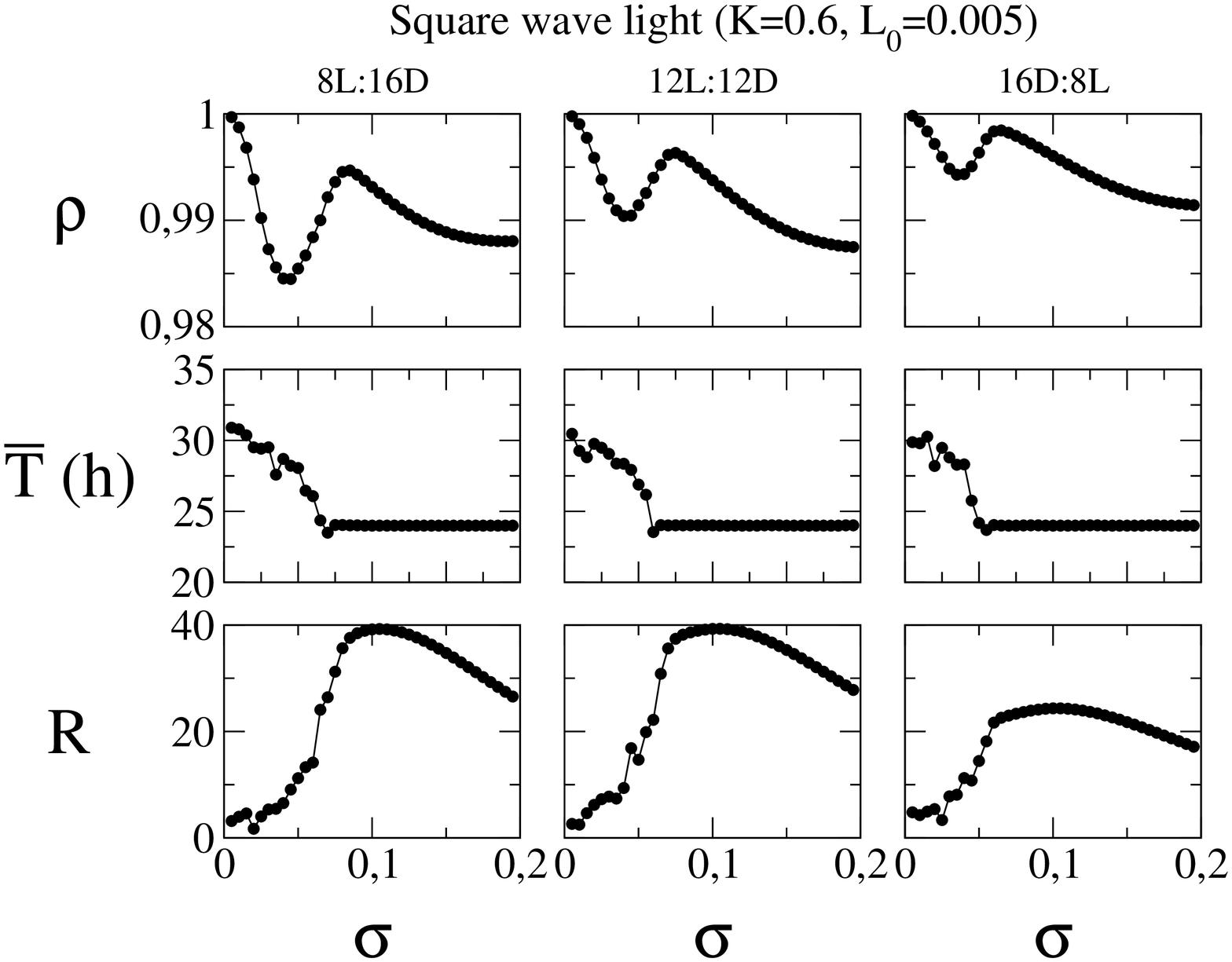}
  \end{center}
  \caption{Characterization of synchrony for a light signal of square wave form. The synchrony parameter $\rho$, the mean $\bar T$ of the individual periods $T_i$ and the response order parameter $R$ (from top to bottom) measured for square wave stimuli of various day lengths ($8h$, $12h$ and $16h$, from left to right).}
  \label{fig:orderParametersRect}
\end{figure}

Fig. \ref{fig:contourRho} shows colour plots of the
parameter of synchrony $\rho$ as a function of the diversity
$\sigma$ and the light intensity $L_0$, for different values of
the coupling strength $K$. High values of the light intensity
$L_0$ favor interneuronal synchrony. Also in agreement with its
intuitive disordering role, high neuronal diversity leads to a
low synchrony parameter $\rho$ in several parts of the
diagrams. However, there is a region of values of
$L_0\in[0,L_{max}]$ for which there is a non-monotonous
dependence of the synchrony order parameter with respect to the
diversity. This can be seen more clearly in panel (a) of
Fig.~\ref{fig:orderParameters} where we plot $\rho$ as a
function of diversity $\sigma$ for fixed values of $K=0.6$ and
$L_0=0.005$. $\rho$ first decreases by increasing $\sigma$
within the interval $0\leq\sigma\leq 0.05,$ but then it
develops a maximum. The range of values of $L_0$ for which this
non-monotonous behavior is observed depends on the coupling
constant $K$: the larger $K$, the larger $L_{max}$.

\begin{figure}
  \begin{center}
    \includegraphics[width=1.\textwidth]{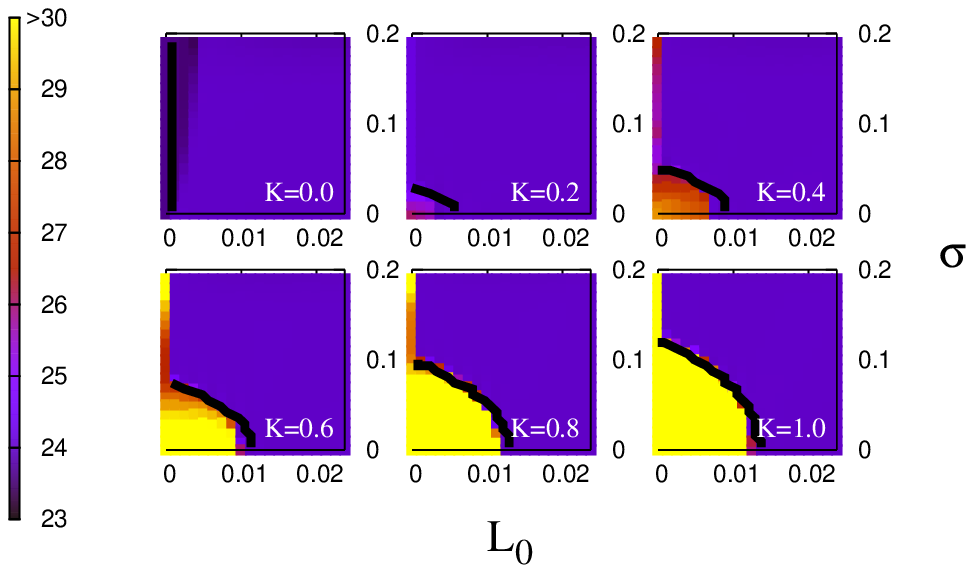}
  \end{center}
  \caption{Colour level plots of the mean of the individual periods $\bar{T}$ of 1000 neurons under forcing at 24h cycle for the system of Eqs. (\ref{eq:Gonze1}--\ref{eq:Gonze5}). High light intensity $L_0$ and high diversity $\sigma$ assures entrainment of oscillations to external frequency (blue region). Increasing coupling enlarges the region (yellow) of oscillations at a period larger than that of the driving force.}
  \label{fig:contourPeriods}
\end{figure}

As stated before, the fact that neurons synchronize amongst
themselves does not mean that they synchronize to the forcing
by light. To study this point, we have computed the individual
periods $T_i$, $i=1,\dots,N$, of the oscillators in the
ensemble. In those cases in which the concentrations do not oscillate with exact periodicity, we still define the period as the average time between maxima of the dynamical variables.
In Fig. \ref{fig:contourPeriods} we plot the mean
value $\bar T=\frac{1}{N}\sum_{i=1}^NT_i$ as a function of
$\sigma$ and $L_0$ for different values of $K$. As the
dispersion in $T_i$ is small, it turns out that $\bar T$
is close to the period of the average
variable ${\bf X}(t)$.

\begin{figure}
    \begin{center}
      \includegraphics[width=1.\textwidth]{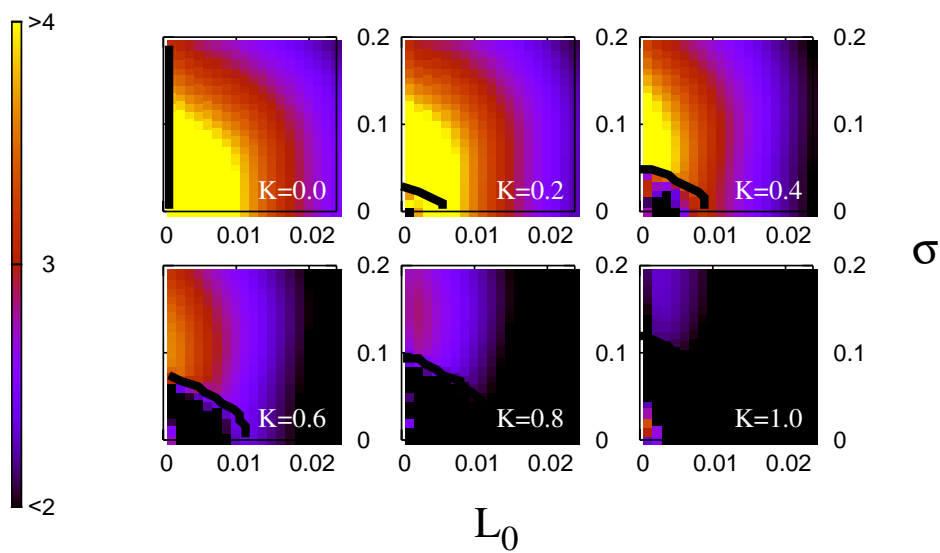}
    \end{center}
    \caption{Colour plots of the spectral amplification factor $R$ as defined in Eq.~(\ref{eq:R}) in logarithmic scale. Too high light intensity $L_0$ and too much diversity $\sigma$ lower the response of $X$ to the external frequency. This also happens for low light and low diversity (where the neurons oscillate at a frequency $\neq 24$ h, see Fig. \ref{fig:contourPeriods}). Between both limits $R$ passes a local maximum.}
    \label{fig:contourR}
\end{figure}

Although, by construction, individual neurons have periods that fluctuate around  $T=23.5$~h, it turns out that the period of the resulting synchronized oscillations that occur in the unforced but coupled ($L_0=0,\,K>0$) case, increases with increasing coupling $K$. For example, $\bar T\approx 30$~h for $K=0.6$, mostly independent of the value of $\sigma$. As the forcing sets in, at low values of the coupling strength, the mean period is now $\bar T=24$~h for all values of $L_0$ and $\sigma$. As the coupling between neurons increases, larger values of $L_0$ and/or $\sigma$ are needed in order for the mean period to coincide with that of the external forcing. An important feature that emerges from these plots is that for low light intensity it is possible to achieve a mean period of $24$~h by increasing the neuronal diversity. For example, in the areas at the left of the different panels of Fig.\ref{fig:contourPeriods}, or in panel (b) of Fig.\ref{fig:orderParameters} corresponding to the case $K=0.6$, while identical neurons have periods of $\approx 30$~h, increasing $\sigma$ induces an adjustment of the period to $24$~h. The transition between $\bar T=24$~h and $\bar T\ne 24$~h is rather sharp, specially for large $K$. This is a clear manifestation that diversity indeed is able to improve the response to the external forcing. The same conclusion about the constructive role of diversity can be reached by looking at the measure of response $R$ (see Figs. \ref{fig:contourR} and \ref{fig:orderParameters}(c)). These figures show that there is a region in parameter space in which  the system response to the periodic light forcing displays a maximum value as a function of diversity $\sigma$. This indicates that it is possible to improve neuronal synchronization to the daily-varying light input by taking $\sigma$ close to an optimal value. Too small or too large diversity will not yield an optimal response. This is a clear manifestation that diversity indeed is able to improve the response to the external forcing.

A complementary perspective on this constructive role of diversity is attained looking at spectral amplification factor, $R$, from Eq.~(\ref{eq:R}). This is a normalized measure of the amplitude of the oscillation of the neuronal system at the frequency of the daily forcing. Figures \ref{fig:contourR} and \ref{fig:orderParameters}(c) show that there is a region in parameter space in which the system response to the periodic light forcing displays a maximum value as a function of diversity $\sigma$. In fact this maximum is very large as compared with the $R$ value at zero diversity, so that one can say that one of the most noticeable effects of a non-vanishing neuronal diversity is to give the system the capacity to respond efficiently to the 24h forcing in situations of small or no response at this frequency in the absence of diversity (the non-diverse neuronal ensemble could be oscillating at a different frequency, as revealed by high values of $\rho$). In summary, it is possible to largely improve neuronal synchronization to the daily-varying light input by taking $\sigma$ close to an optimal value. Too small or too large diversity will not yield an optimal response at this frequency, although the response is generally larger than for zero diversity.

An external signal of square wave form and with different day lengths lead to similar results. As can be seen in figure~\ref{fig:orderParametersRect} the response $R$ to the external signal passes through a maximum at a intermediate value of diversity. The mean period and the synchrony parameter behave as in the case with a pure sinusoidal as the driving force. Furthermore, the qualitative result is independent of the chosen day length. 

\subsection{Diversity-induced oscillator death}
\label{sec:Discussion}

\begin{figure}
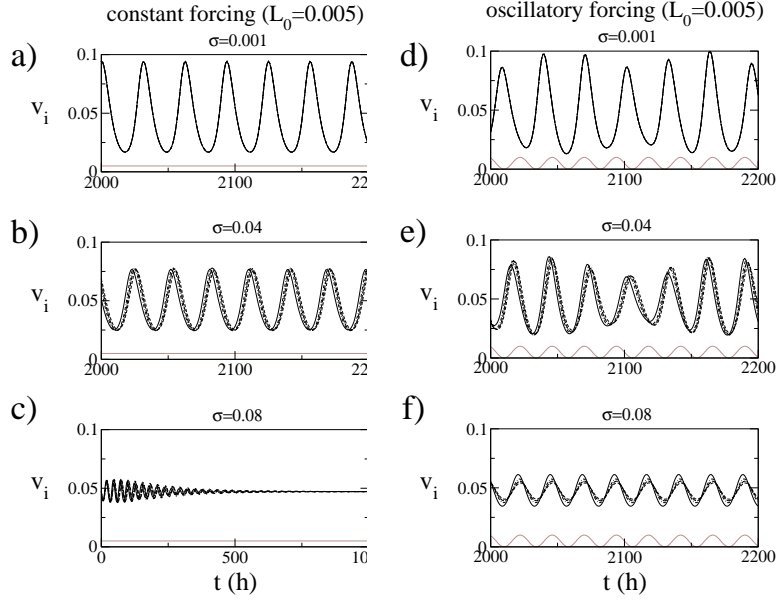

    \begin{center}
     \includegraphics[width=0.4\textwidth]{fig_6a.eps}
     \includegraphics[width=0.4\textwidth]{fig_6b.eps}
      \end{center}
    \caption{Figures (a), (b) and (c) represent the time-dependent amplitude of the $V_i$ variable for a few selected neurons in the presence of constant light and increasing $\sigma$, while Figures (d), (e) and (f) represent the amplitude of the same neurons with sinusoidal light and increasing $\sigma$. The thin line on the bottom of the graphs is the external light signal. $K=0.6$}
    \label{fig:concEvolution}
\end{figure}

Why does an increase in the diversity of the oscillators lead
to an improved response to the external forcing? We argue that
the main effect of the increase of the diversity is to take the
oscillators into a regime of oscillator death
\citep{Ermentrout1990219,MS:1990} in which they can be easily entrained by
the varying part of the forcing. To understand this mechanism
we first split the forcing into a constant (the mean) and a
time varying part: $L(t)=\displaystyle
\frac{L_0}{2}+\frac{L_0}{2}\sin(\omega t)$. Taking only the
constant part, $L(t)=\displaystyle \frac{L_0}{2}$,
Figs.~\ref{fig:concEvolution}(a)--(c) show that the oscillators
go from self-sustained oscillations to oscillator death, i.e.
the amplitude of the self-sustained oscillations decreases, as
$\sigma$ increases. Once oscillators are damped, they would
respond quasi-linearly to periodic forcing, at least if this
forcing is not too large, and linear oscillators always become
synchronized to the external forcing, independently of their
internal frequency. This is consistent with what is seen in
figures~\ref{fig:concEvolution}(d)--(f), where the neurons in
the case of low heterogeneity oscillate synchronously with each
other, but their common period is larger than the one of the
light forcing. Only when diversity brings the neurons to
oscillator death can all of them be entrained to the period of
the forcing signal. The mechanism is related to the one
discussed by \citet{DidierGonze07012005} and \citet{BJ07}, but here we stress that neuron
heterogeneity, as opposed to internal neuron parameters and
couplings, is enough to damp the collective neuron oscillations
and bring the system to a non-oscillating state where it can be
more easily entrained.  It is interesting to note that it has been
shown experimentally for fruitflies that only a subset of the pacemaker neurons
sustain cyclic gene expression after changing the laboratory light conditions 
to constant darkness, whereas the oscillations of the other pacemaker neurons are damped out~\citep{Veleri2003}. Although this does not reveal the mechanism by which the oscillations die out it suggests that some of the circadian oscillators do indeed work in the damped regime, at least in {\sl Drosophila}.

\begin{figure}[ht]
  \begin{center}
    \includegraphics[width=1.\textwidth]{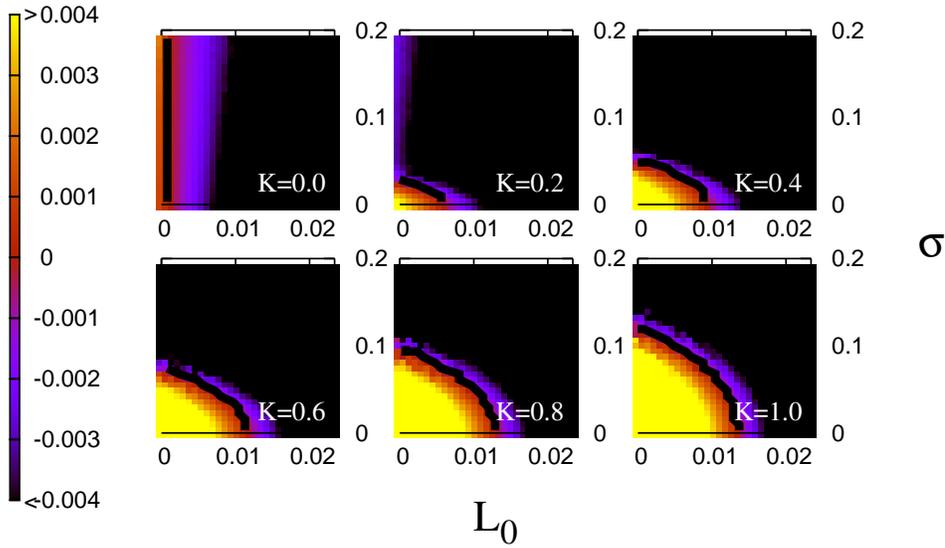}
  \end{center}
  \caption{Colour plots of the maximum real part of the average eigenvalues of system of Eqs. (\ref{eq:Gonze1}--\ref{eq:Gonze5}), as a function of $\sigma$ and $L_0$, at different values of $K$. Increasing $\sigma$ or increasing $L_0$ changes this quantity from positive to negative, i.e. transforms the self-sustained neurons into damped neurons by stabilizing their constant concentrations fixed points. Rising the coupling enlarges the region of self-sustained oscillations. Averaged from 10 realizations of heterogeneity in 200 neurons.}
  \label{fig:contourEVS}
\end{figure}

An alternative way of checking this mechanism based on {\sl
diversity-induced oscillator death} is by analyzing the
stability of the steady state of the system of Eqs.
(\ref{eq:Gonze1}--\ref{eq:Gonze5}) when considering a constant
forcing $L(t)=\displaystyle \frac{L_0}{2}$. The numerical
calculation of the fixed point of the dynamics is greatly
simplified by the fact that the concentrations of the
biochemical variables are the same for each one of the $N$
neurons irrespectively of their specific value of $\tau_i$. The
system (\ref{eq:Gonze1}--\ref{eq:Gonze5}) is linearized around
this steady state and the eigenvalues of the stability matrix
computed for several realizations of diversity parameters
${\tau_i}$. In each case, the positive or negative character of
the real part of the eigenvalue with the largest real part
indicates the instability or stability, respectively, of the
fixed point solution. In Fig.~\ref{fig:contourEVS} we plot the
mean of that maximum real part of the eigenvalues averaged over
various realizations of the time scales ${\tau_i}$, for $N=200$
coupled neurons, as a function of $L_0$ and $\sigma$, and
different values of the coupling $K$ (see also panel (d) in
Fig.\ref{fig:orderParameters}). In every diagram we can see
that low diversity or low forcing yield an unstable steady
state (yellow region). This is where self-sustained
oscillations are observed. A thick black line in the contour
plots indicates a zero real part. The relevance of this line
separating positive from negative maximum average eigenvalues
is more apparent when we note that it also delimits regions of
interest in Figs.\ref{fig:contourRho}, \ref{fig:contourPeriods}
and \ref{fig:contourR}.

In summary, increasing the diversity or the (constant) forcing
term decreases (on average) the maximum eigenvalue of the
coupled system and thus a Hopf bifurcation can be crossed
backwards, so that self-oscillations disappear. When applying
the periodic external forcing on the system formed by
self-sustained neurons, coherence with the external frequency
is difficult to achieve because there is the competing effect
of mutual neuron synchronization to a different frequency.
However, when the periodic external forcing is applied on the
system of damped neurons, they all synchronize to the external
forcing, and thus with each other since this is the only
dynamical regime available to forced damped oscillators (if
forcing is not too strong to excite further resonances).
Increased coupling strength increases the range of unentrained
self-oscillations.

Oscillator death by diversity is not particular to this system.
In~\citep{MS:1990} the authors analyze a large system of limit
cycle-oscillators with mean field coupling and randomly
distributed frequencies. They proved that when the coupling is
sufficiently strong and the distribution of frequencies has a
sufficiently large variance, the system undergoes ``amplitude
death''. In their approach the oscillators pull each other off
their limit cycles, which is translated into a stable
equilibrium point for the coupled system. Thus, this mechanism
suggests that the quenched noise we introduced in the system
``pushes apart" the limit cycles of the different neurons, so
that their competition enlarges the range of parameters where
fixed point behavior is stable.

A qualitative argument explaining the diversity-induced oscillator death in our system of coupled neurons goes as follows: We know from \citet{DidierGonze07012005} that a single oscillator can switch from a limit cycle to a stable steady state by adding a constant mean field (the term containing $F$ in (\ref{eq:Gonze1}) but with time-independent $F$) of sufficient strength to Eq.~(\ref{eq:Goodwin1}). A constant light forcing term has the same effect (see the zero coupling case in fig.~\ref{fig:contourEVS}). Furthermore we have observed that the amplitude of the oscillations decreases with rising diversity (compare figs.~\ref{fig:concEvolution}), but the mean does not change. In a system with low diversity we have large oscillations of $F$ around that mean value. If this value, taken as a constant, determines a stable steady state, then we argue that the large oscillations lead the system into unstable regions, whereas, by increasing $\sigma$ the amplitude is decreased and the concentrations do not leave neighbourhood of the stable fixed point, thus finding themselves damped all the time. This is a possible mechanism for the {\sl diversity-induced oscillator death} phenomenon.

\section{Concluding Remarks}
\label{sec:Conclusion}

In this work we have analyzed the role of diversity in favoring the entrainment of a system of coupled circadian oscillators. We introduce non-negligible heterogeneity in the periods of all neurons in the form of quenched noise. This is achieved by rescaling the individual neuronal periods by a scaling factor drawn from a normal distribution. The system response to the light-dark cycle periodicity is studied as a function of the interneuronal coupling strength, external forcing amplitude and neuronal heterogeneity.

Most of the cases of order induced by heterogeneity or noise carried out so far \citep{Gammaitoni1998,HM,TMTG:2006,tessone2007a,toral2008a,PK:1997,Ullner2009}, emphasize the fact the diversity directly improves oscillator synchronization. In our case the mechanism is rather different. Diversity does not improve system synchronization directly. This is achieved indirectly, by a leading first to a diversity-induced stabilization of the fixed points of the neurons forming the system. Once steady concentrations are asymptotically stable, it is much better entrainable by the external forcing, so that the damped neurons adapt easily to the external forcing (and then, in addition, they appear as synchronized between them). The synchronization arises therefore not as a result of a direct diversity-induced neuronal synchronization but indirectly, as a result of the diversity-induced oscillator death. Our results indicate therefore that the right amount of heterogeneity helps the extended system to respond globally in a more coherent way to the external forcing.  In addition to the robustness of the results against use of non-sinusoidal forcing we have checked that resonances in the responses to the external forcing and matching of the circadian period to the external forcing appear in more complex models, such as the 10-variable model of~\citep{BJ07} with diversity in the time scales $\tau_i$, or the 4-variable model of~\citep{DidierGonze07012005} with heterogeneity in all the reaction rate parameters $\nu_i$. We expect that a similar behavior will be found in models of non-mammalian clocks like those of {\it Drosophila}~\citep{Smolen2004}, {\it Arabidopsis}~\citep{Locke2005}, {\it Neurospora}~\citep{Heintzen2007} or Cyanobacteria~\citep{Dong2008}.

Of course, it is an open question whether the observed diversity in the periods of the neurons of the SCN has been tuned by evolution in order to display a maximum response to the $24$~h dark-light natural cycle. A detailed experimental check of our predictions would require to be able to vary the amount of diversity. In this sense we suggest the possibility of using chimeric organisms~\citep{LowZeddies2001} to introduce diversity in a controlled way.

\section*{Acknowledgements} 
We acknowledge financial support from
the EU NoE BioSim, contract LSHB-CT-2004-005137, and MEC
(Spain) and FEDER (EU) through project FIS2007-60327. NK is
supported by a grant from the Govern Balear.


\label{lastpage}
\end{document}